\documentclass[aps,prb,reprint,superscriptaddress,showpacs]{revtex4-1}
\usepackage{amsmath}
\usepackage[pdftex]{graphicx}
\usepackage{braket}
\usepackage{subfigure}
\usepackage{natbib}

\begin{document}
\DeclareGraphicsExtensions{.png}

\title{Autler-Townes splitting in a three-dimensional transmon superconducting qubit}
\date{\today}

\author{S. Novikov}
\email{snovikov@umd.edu}
\affiliation{Department of Physics, University of Maryland, College Park, Maryland 20742, USA}
\affiliation{Laboratory for Physical Sciences, College Park, Maryland 20740, USA}
\author{J. E. Robinson}
\affiliation{Laboratory for Physical Sciences, College Park, Maryland 20740, USA}
\author{Z. K. Keane}
\affiliation{Department of Physics, University of Maryland, College Park, Maryland 20742, USA}
\affiliation{Laboratory for Physical Sciences, College Park, Maryland 20740, USA}
\author{B. Suri}
\affiliation{Department of Physics, University of Maryland, College Park, Maryland 20742, USA}
\affiliation{Laboratory for Physical Sciences, College Park, Maryland 20740, USA}
\author{F. C. Wellstood}
\affiliation{Department of Physics, University of Maryland, College Park, Maryland 20742, USA}
\affiliation{Joint Quantum Institute and Center for Nanophysics and Advanced Materials, College Park, Maryland 20742, USA}
\author{B. S. Palmer}
\affiliation{Laboratory for Physical Sciences, College Park, Maryland 20740, USA}


\begin{abstract}
We have observed the Autler-Townes doublet in a superconducting Al/AlO$_{\mbox{x}}$/Al transmon qubit that acts as an artificial atom embedded in a three-dimensional Cu microwave cavity at a temperature of 22 mK. Using pulsed microwave spectroscopy, the three lowest transmon levels are isolated, eliminating unwanted effects of higher qubit modes and cavity modes. The long coherence time ($\sim 40\; \mu s$) of the transmon enables us to observe a clear Autler-Townes splitting at drive amplitudes much smaller than the transmon level anharmonicity (177 MHz). Three-level density matrix simulations with no free parameters provide excellent fits to the data. At maximum separation, the fidelity of a dark state achieved in this experiment is estimated to be $99.6-99.9\%$.
\end{abstract}

\maketitle
The Autler-Townes (AT) effect involves a three-level quantum system interacting with an applied coupling drive field \cite{Autler1955, cohen-tannoudji_atom-photon_1998} that is nearly resonant with two of the levels. For a sufficiently strong coupling field, one of the transitions will split into a doublet, which can be probed by a weak second tone. The effect is an example of electromagnetic dressing of quantum states, and it has been proposed as a basis for fast, high \textsc{on/off} ratio microwave routers \cite{Li2012, Hoi2013} for quantum computation. Furthermore, the AT effect is closely related to electromagnetically-induced transparency \cite{Boller1991} (EIT) and quantum effects such as slow light.\cite{Turukhin2001} Observing EIT poses more stringent requirements on the coherence of the system, and although EIT has been shown in atomic systems, \cite{Field1991} it has not been decisively demonstrated with superconducting qubits. \cite{Anisimov2011} EIT in superconducting systems has been proposed as a sensitive probe of decoherence.\cite{Murali2004}

The AT effect has been studied in atomic,\cite{Picque1976, Cahuzac1976, Delsart1976} molecular systems,\cite{Tamarat1995} quantum dots,\cite{Xu2007} and superconducting qubits. Groups studying the effect in superconducting qubits have employed transmon levels with continuous-tone cavity readout,\cite{Baur2009} phase qubit levels with tunneling readout,\cite{Sillanpaa2009, Kelly2010} flux \cite{Abdumalikov2010} and transmon \cite{Hoi2013} qubit levels coupled directly to a transmission line.

In this article, we present experimental measurements of the AT splitting in a 3D transmon superconducting qubit.\citep{Paik2011} In contrast to previous studies involving transmons,\cite{Baur2009} which had to include transmon-cavity effects of the Jaynes-Cummings Hamiltonian, we isolate three lowest transmon levels by using pulsed spectroscopy. Our method eliminates the need to re-tune microwave drives to account for power-dependent dispersive shifts,\citep{Baur2009} and we achieve a large signal-to-noise ratio by using a qubit-induced-nonlinearity readout.\cite{Reed2010, Bishop2010, Boissonneault2010} Previous experiments using superconducting qubits \cite{Sillanpaa2009} employed coupling drives that were large compared to the energy level anharmonicity to compensate for relatively short coherence times. At such strong drives, multi-photon transitions are possible, and accurate modeling of the system requires a Hilbert space of more than three levels. In contrast, our device possesses long enough coherence times to observe the AT doublet even at low drive amplitudes, and we show that our data is well-explained using a three-level density matrix with no free parameters. 

\begin{figure}
	\includegraphics[width=8.6cm]{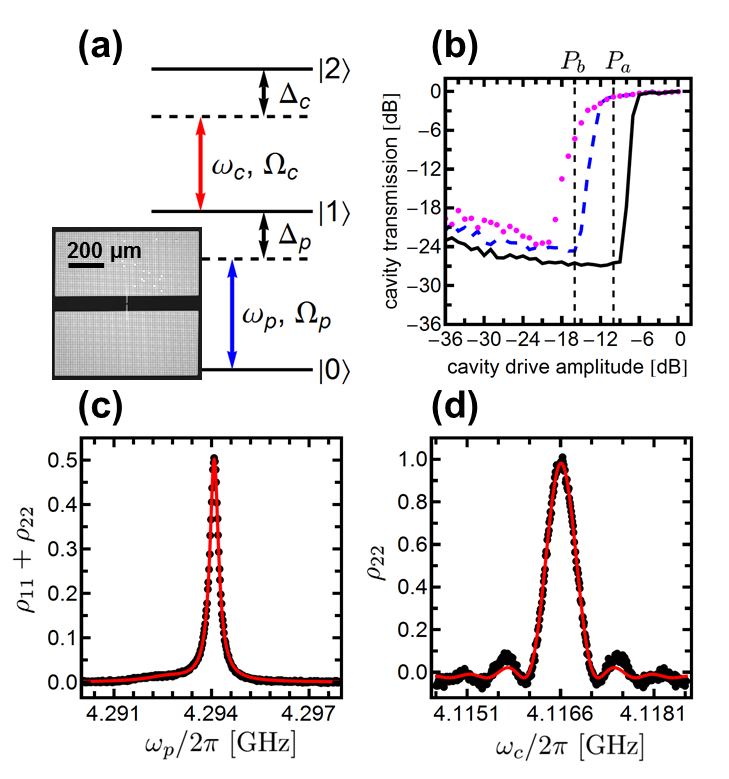}
	\caption{\label{fig:levels}(Color online). (a) Energy level diagram for the transmon indicating the levels and microwave drives. Inset: optical micrograph of transmon showing Al pads (light grey) on sapphire (dark grey). (b) Cavity transmission at $\omega_{cav}/2\pi = 7.1585\mbox{ GHz}$ after the transmon is prepared in state $\ket{0}$ (black line), $\ket{1}$ (blue dash) or $\ket{2}$ (magenta dots). Vertical dashed lines indicate drive powers $P_a$ used in the AT experiment to achieve the readout proportional to $\rho_{11}+\rho_{22}$, and $P_b$ to readout $\rho_{22}$ for $\omega_{12}$ calibration. (c) Spectroscopic measurements (black dots) and fit (red line) of $\omega_{01}$. Note small shoulder at $4.292$ GHz near the 0-to-1 transition peak at $4.294$ GHz. (d) Spectroscopic measurements (black dots) and fit (red line) of $\omega_{12}$.}
\end{figure}

We model the system by considering just the ground, first, and second excited states of the transmon: $\ket{0}$, $\ket{1}$, and $\ket{2}$. These are separated by two transition frequencies: $\omega_{01}$ and $\omega_{12} \equiv \omega_{02} - \omega_{01} = \omega_{01} + \alpha$, where $\omega_{ij} \equiv \omega_j - \omega_i$, and $\alpha$ is the level anharmonicity. Two microwave drives, the probe and the coupler, are applied at $\omega_p = \omega_{01} + \Delta_p$ and $\omega_c = \omega_{12} + \Delta_c$ [see Fig. \ref{fig:levels}(a)]. Their amplitudes $\Omega_p$ and $\Omega_c$ determine the Rabi oscillation frequencies of the $\omega_{01}$ and $\omega_{12}$ transitions, respectively.  In the frame co-rotating with the drives, the system Hamiltonian is
\begin{eqnarray}\label{eq:ham}
	\mathcal{H} = &-&\hbar\Delta_p \ket{1}\bra{1} - \hbar(\Delta_p + \Delta_c) \ket{2}\bra{2} \notag \\&+& \Bigg(\hbar\frac{\Omega_{p}}{2}\ket{1}\bra{0} + \hbar\frac{\Omega_{c}}{2}\ket{2}\bra{1} + \mbox{H.c.}\Bigg)
\end{eqnarray}
Dissipation and dephasing are included via the Kossakowski-Lindblad \cite{Kossakowski1972, Lindblad1976} master equation for the density matrix $\rho$
\begin{equation}\label{eq:master}
\frac{d\rho}{dt} = - \frac{i}{\hbar} [\mathcal{H},\rho] + \sum_j \Big[ \mathcal{L}_j\rho\mathcal{L}_j^\dagger - \frac{1}{2}\Big( \rho \mathcal{L}_j^\dagger \mathcal{L}_j + \mathcal{L}_j^\dagger \mathcal{L}_j \rho \Big)\Big],
\end{equation}
where $\mathcal{L}_j$ is the Lindblad operator describing decoherence of the system through a particular channel $j$. We solve Eq. \ref{eq:master} numerically in steady-state to obtain the theoretical description of our data.\cite{Li2011, supplemental} As expected for the AT effect, one finds a splitting of the $\omega_{12}$ transition that increases with increasing coupler power.

Our device consists of a transmon \cite{Koch2007} embedded in a 3D microwave cavity.\cite{Paik2011, Rigetti2012} The transmon is a superconducting artificial atom whose Hamiltonian resembles that of an anharmonic oscillator. It is formed by the non-linear inductance of a Josephson junction shunted by a capacitor. The transmon [see inset Fig. \ref{fig:levels}(a)] is made via standard e-beam lithography, double-angle evaporation,\cite{Dolan1977} and lift-off procedures. It has a single Al/AlO$_{\mbox{x}}$/Al Josephson junction capacitively shunted by two 375$\times$800 $\mu$m Al pads on a sapphire substrate. The tunneling energy of the junction is $E_J/h = 16.5\mbox{ GHz}$, and the pads lower the charging energy to $E_C/h = 177\mbox{ MHz}$. The pads are fabricated as a mesh of 2.5 $\mu$m wide lines placed every 10 $\mu$m in both directions to enhance expulsion of any external magnetic fields and trap any magnetic vortices already present. The shunting pads also form a dipole antenna which couples the qubit to the cavity with strength $g/2\pi = 151\mbox{ MHz}$. The 3D cavity is a rectangular box made from oxygen-free-high-conductivity Cu, with the fundamental TE$_{101}$ mode at $\omega_{cav}/2\pi = 7.1585\mbox{ GHz}$. This mode is used for qubit readout, with loss limited by the internal quality factor $Q_i = 18,000$. The cavity is probed in transmission, with the output connector coupled much more strongly ($Q_e^{out} = 30,000$) than the input connector ($Q_e^{in} = 120,000$). The cavity is mounted on the mixing chamber of a Leiden Cryogenics CF-450 dilution refrigerator at $T = $ 22 mK. The microwave lines to the cavity are heavily attenuated, filtered and isolated to protect the device from extrinsic noise. The output signal from the cavity is passed to a high-electron-mobility transistor amplifier at the 3 K stage, and then further amplified, mixed down, and digitized at room temperature.\cite{supplemental}

Three microwave drives are used: cavity, probe, and coupler. The cavity is turned on at time $t=0$ for 5 $\mu$s to record the initial (ground) state of the system, and then again at $t=290\;\mu$s to read out the final state of the system. Within the 290 $\mu$s window between readout pulses, transmon control microwaves (either probe or coupler, or both) are applied. The whole sequence is repeated every 600 $\mu$s. The cavity is used solely for the readout, and does not participate in the Autler-Townes manifold. 

Measurement of the transmon state is achieved with a high signal-to-noise ratio by using the Jaynes-Cummings non-linearity readout.\cite{Reed2010, Bishop2010, Boissonneault2010} The cavity pulses are applied at the bare cavity frequency $\omega_{cav}/2\pi = 7.1585\mbox{ GHz}$ and with an amplitude that provides maximum contrast between the ground and excited states [Fig. \ref{fig:levels}(b)]. At amplitude $P_a$ the cavity does not discriminate between the transmon being in $\ket{1}$ or $\ket{2}$, while at $P_b$ it is mostly sensitive to $\ket{2}$. The signal from the non-linearity readout at $P_a$ is proportional to the sum total of the first and second excited state probabilities, $\rho_{11}$ and $\rho_{22}$, and is used to obtain the 0-to-1 and AT data. The signal at $P_b$ is used for the characterization of 1-to-2 transition only. 

We can determine the parameters in Eqs. \ref{eq:ham} and \ref{eq:master} from a set of measurements on $\omega_{01}$ and $\omega_{12}$ as follows. We characterized the $\omega_{01}$	 transition by applying just the probe and cavity readout tones. By pulsing the probe we find the relaxation time $T_1 = 39\;\mu s$, limited by internal loss. The measured Ramsey decay time $T_2^* = 51\;\mu s$ is less than the relaxation-limited value of $2T_1$ due to additional dephasing. This $T_2^*$ can be used to place a bound of less than $0.02$ thermal photons in the cavity.\cite{Sears2012} From $T_1$ and $T_2^*$ we obtain relaxation (denoted by $\Gamma_{ij}$ for $\ket{i}\rightarrow\ket{j}$ process) and dephasing (denoted by $\gamma_i$ for state $\ket{i}$) rates of $\Gamma_{10} = 1/T_1 = 26\times 10^{3}\mbox{ s}^{-1}$, $\gamma_2 = \gamma_1 = 1/T_2^* - 1/2T_1 = 6.6\times 10^{3}\mbox{ s}^{-1}$. We set $\Gamma_{21} = 1.41 \times 26\times 10^{3}\mbox{ s}^{-1}$ and $\Gamma_{20} = 0$ based on the ratio of transmon transition matrix elements.\cite{Koch2007} We assume negligible upward rates in the system, and set $\Gamma_{ij} = 0$ for all $i<j$.

\begin{figure}
	\includegraphics[width=8.6cm]{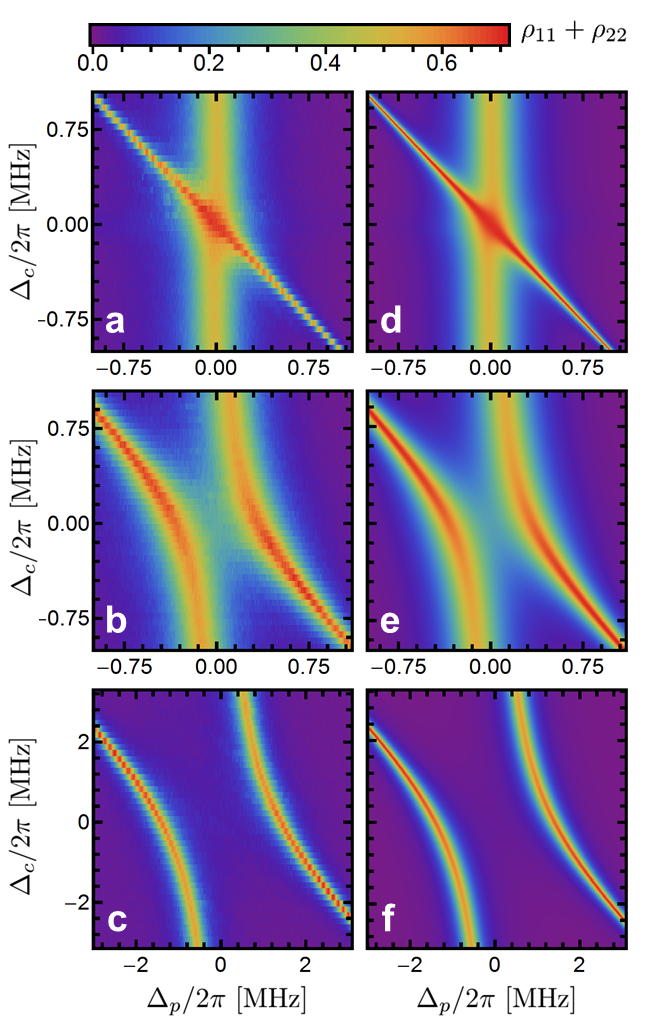}
	\caption{\label{fig:SplittingData}(Color). (a-c) data and (d-f) simulations of the Autler-Townes splitting for several coupler powers. Coupler strengths are: (a) and (d) 0.177 MHz, (b) and (e) 0.707 MHz, (c) and (f) 2.82 MHz. To account for larger peak separation, the scale is increased on the bottom row of plots.}
\end{figure}
\begin{figure}	
	\includegraphics[width=8.6cm]{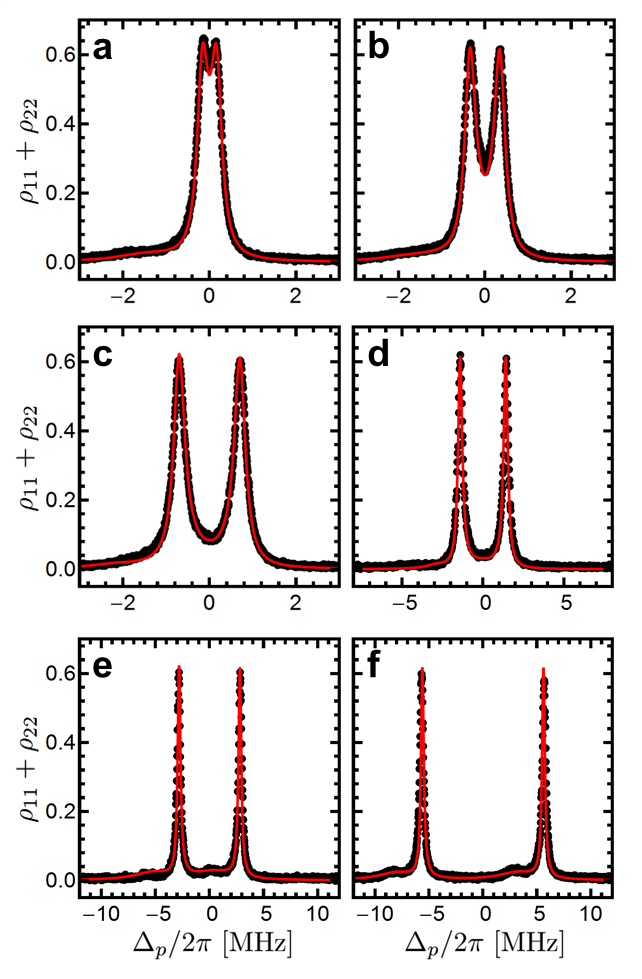}
	\caption{\label{fig:splittingVSpower}(Color online). Data (black dots) and simulation (red line) of the Autler-Townes doublet at $\Delta_c = 0$. Coupler strengths are: (a) 0.354 MHz, (b) 0.707 MHz, (c) 1.41 MHz, (d) 2.82 MHz, (e) 5.63 MHz, (f) 11.2 MHz.}
\end{figure}
At low probe powers, we observe a small shoulder on the left-hand side of the $\omega_{01}$ peak [see Fig. \ref{fig:levels}(c)], which we attribute to a fluctuator affecting the transmon. Similar fluctuators, possibly due to a microscopic defect in or near the junction, have been studied in other superconducting qubits.\cite{Martinis2005, Constantin2007, Zaretskey2013} Apart from the slight background, the fluctuator does not affect the system in any way. We fit the data using the steady-state solution to Eq. \ref{eq:master}, with an additional Lorentzian to account for the fluctuator background [Fig. \ref{fig:levels}(c)]. With decoherence rates determined independently, we extract best fit values $\Omega_p/2\pi = 186\mbox{ kHz}$, $\omega_{01}/2\pi = 4.294\;085\mbox{ GHz}$, and the position, width and amplitude of the Lorentzian background.

In order to characterize the $\omega_{12}$ transition, we perform a $\pi$-pulse at $\omega_{01}$ with the probe tone, followed by a $\pi$-pulse near $\omega_{12}$ with the coupler. We measure the population of $\ket{2}$ alone at a cavity power $P_b \approx P_a - 10\mbox{ dB}$ that provides contrast only when $\ket{2}$ is excited [Fig. \ref{fig:levels}(b)]. The spectroscopic peak [Fig. \ref{fig:levels}(d)] is fit to obtain $\omega_{12}/2\pi = 4.116\;609\mbox{ GHz}$. To calibrate $\Omega_c$ for the AT experiment, the Rabi frequency of the $\ket{1}\leftrightarrow\ket{2}$ transition as a function of coupler amplitude is measured by replacing the $\pi$-pulse on the coupler with a variable-length pulse at $\Delta_c = 0$.

Finally, we calibrate the probability scale by performing Rabi oscillations on $\omega_{01}$. We fit the data, and set the amplitude of the fit exponentially decaying sine function to unity. This calibrates $\rho_{11}$, and, with the readout at cavity power $P_a$ being equally sensitive to $\rho_{11}$ and $\rho_{22}$, also calibrates $\rho_{11}+\rho_{22}$.

For the Autler-Townes experiment, the probe and the coupler are turned on for 280 $\mu$s between the two cavity readout pulses. Being much longer than any coherence times in the system, this probe and coupler pulse length ensures the system has achieved a steady state before the measurement. Sweeping both $\Delta_p$ and $\Delta_c$ around zero, and measuring $\rho_{11}+\rho_{22}$, we observe emergence of the Autler-Townes doublet as $\Omega_c$ is increased [see Fig. \ref{fig:SplittingData}]. At the relatively low coupling drive of $\Omega_c/2\pi = 0.177\mbox{ MHz}$ [Fig. \ref{fig:SplittingData}(a)] we see a crossing of $\omega_{01}$ (vertical band) with the two-photon sideband excitation of $\omega_{02}$ (diagonal streak). As the coupler strength is increased four-fold, $\omega_{01}$ becomes dressed by the coupler photons and shows the emergence of an anti-crossing at zero detuning [Fig. \ref{fig:SplittingData}(b)]. Increasing $\Omega_c$ another four-fold results in a completely separated splitting [Fig. \ref{fig:SplittingData}(c)]. Figures \ref{fig:SplittingData}(d-e) show the corresponding simulations found by solving Eq. \ref{eq:master} with no fitting parameters. We find excellent agreement with the data.

To observe a well-separated AT doublet, we must apply sufficiently strong coupler tone while keeping excitations to a three-level manifold. The anharmonicity of the device, $\alpha \equiv \omega_{01} - \omega_{12} = E_C/\hbar = 2\pi \times 177\mbox{ MHz}$, sets an upper limit for the strengths of the drives that can be used. The proximity of the $\ket{0}\!\leftrightarrow\!\ket{2}$ two-photon transition at $\omega_{02}/2 = \omega_{01}-\alpha/2$ can also, at sufficiently strong drives, interfere with AT signal.\citep{Sillanpaa2009} Although the transitions are power-broadened to $\Gamma/2\pi\approx 350\mbox{ kHz}$ at this probe amplitude, they remain much smaller than $\alpha$. Therefore, we require $\Omega_c \gg \Gamma$, $\Omega_c \gg \Omega_p$ for a well-separated AT doublet, as well as the  $\Omega_c \ll \alpha$ to restrict the Hilbert space to the three lowest levels.

For coupler detuning $\Delta_c = 0$, the splitting is symmetric around probe detuning $\Delta_p = 0$. As Fig. \ref{fig:splittingVSpower} shows, we see excellent agreement between the data and the density matrix simulation with all parameters independently determined, and an additional Lorentzian added at $\pm \Omega_c/2$ to account for the small background due to the aforementioned fluctuator.

At $\Delta_c = \Delta_p = 0$, the eigenstates of the system can be written in a simple form:\cite{Li2011}
\begin{eqnarray}
	\ket{D} &=& \mbox{cos}\,\Theta\ket{0} - \mbox{sin}\,\Theta\ket{2} \\
	\ket{+} &=& \frac{1}{\sqrt{2}}[\mbox{sin}\,\Theta\ket{0} + \ket{1} + \mbox{cos}\,\Theta\ket{2}]\\
	\ket{-} &=& \frac{1}{\sqrt{2}}[\mbox{sin}\,\Theta\ket{0} - \ket{1} + \mbox{cos}\,\Theta\ket{2}]
\end{eqnarray}
where the mixing angle $\Theta = \mbox{tan}^{-1}(\Omega_p/\Omega_c)$. State $\ket{D}$ is a dark state with eigenvalue of zero, while states $\ket{\pm}$ correspond to eigenvalues $\pm\sqrt{\Omega_p^2 + \Omega_c^2}$, i.e. separated from the dark state by the generalized Rabi frequency. For large peak separation, the dark state mostly consists of the ground state. In the AT regime the dark state is not achieved by population inversion into $\ket{2}$ but, rather, by having most of the population in $\ket{0}$. Nevertheless, the AT dark state can still be used as the \textsc{off} state in router applications due to vanishing contributions of $\ket{1}$ at large peak separations. The fidelity of the dark state can be defined by \cite{Li2011}

\begin{eqnarray}
\mathcal{F}_{\ket{D}} &=& \sqrt{\bra{D}\rho\ket{D}} \notag \\
&=& \frac{\mbox{cos}\,2\Theta}{2}(\rho_{00}-\rho_{22}) - \frac{\mbox{sin}\,2\Theta}{2}(\rho_{20}+\rho_{02}) \notag \\
&+& \frac{1}{2}(1-\rho_{11}).
\end{eqnarray}

\begin{figure}
	\includegraphics[width=8.6cm]{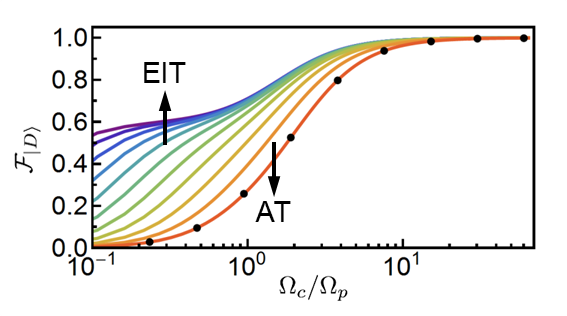}
	\caption{\label{fig:fidelity}(Color online). Dark state fidelity inferred from simulations versus coupler power (black dots), and theoretical fidelity (colored lines) for a system with $\Gamma_{21}' = \Gamma_{21}/2^n$, $n=0,1, \ldots ,9$ (red to violet). A crossover to $\Gamma_{21} \ll \Gamma_{10}$ regime where EIT is possible manifests in the increased fidelity even at small $\Omega_c/\Omega_p$.}
\end{figure}

From the experimental values of $\Omega_p$ and $\Omega_c$, as well as the density matrix elements calculated in the simulations, we can infer dark state fidelities of the observed data (Fig. \ref{fig:fidelity}).
At the two highest coupler powers [$\Omega_c/2\pi = 5.63\mbox{ and }11.2\mbox{ MHz}$, Figs. \ref{fig:splittingVSpower}(e) and \ref{fig:splittingVSpower}(f)] the data starts displaying discrepancies from the simulation due to increased proximity to the $\omega_{02}/2$ transition. This is manifested as the AT doublet being pushed to a higher frequency and not centered at $\Delta_p = 0$. The fluctuator parameters also change slightly. To account for these discrepancies in our calculation of $\mathcal{F}_{\ket{D}}$, we fit both the AT peaks and the background with Lorentzians to determine new probe detuning, $\Delta_p'$, and new background parameters to feed into the simulation. At the maximum separation (29 linewidths), we calculate the dark state fidelity to be $99.6-99.9\%$. We note that even at the largest coupler powers used, the effects of the higher levels are dispersive, manifesting themselves not as additional excitations but as slight frequency shifts of the doublet.

Figure \ref{fig:fidelity} also shows a theoretical prediction for the scaling of fidelity with $\Omega_c/\Omega_p$ if our system were in the EIT regime, which would require $\Gamma_{21} \ll \Gamma_{10}$. We model the EIT regime by replacing $\Gamma_{21}$ by $\Gamma_{21}'=\Gamma_{21}/2^n$ for $n = 0,1 \ldots ,9$ while keeping all other simulation parameters the same. A long-lived $\ket{2}$ makes population trapping in that state possible, opening a narrow EIT window, and resulting in high fidelities even for $\Omega_c/\Omega_p \ll 1$. We believe that the EIT regime can be achieved by engineering the rates of this system.

In summary, we have observed emergence of Autler-Townes splitting in a 3D transmon system by dressing the three lowest levels of the transmon with two drives and reading out the state of the system by pulsing an additional cavity tone. We achieve $99.6-99.9\%$ maximum dark state fidelity at 29 linewidths of separation. Even at the highest coupler powers, the data stands in good agreement with a three-level density matrix simulation. Although we do not have a direct measurement of the dark state fidelity, our technique for determining $\mathcal{F}_{\ket{D}}$ as a function of $\Omega_c/\Omega_p$ using independently measured parameters provides a useful metric for assessing EIT, distinguishing EIT from the Autler-Townes effect, characterizing the \textsc{on/off} ratio, and assessing dark state coherence.

F.C.W. would like to acknowledge support from the Joint Quantum Institute and the State of Maryland through the Center for Nanophysics and Advanced Materials. The authors would like to acknowledge Paul Lett and Eite Tiesinga for their helpful discussions.

\bibliography{Novikov_AT_manuscript_refs}
\end{document}